\documentclass[acmtog]{acmart}
\acmSubmissionID{253}

\usepackage{booktabs} 
\usepackage{resizegather}
\usepackage{algpseudocode}
\usepackage{color, colortbl}
\definecolor{Gray}{gray}{0.85}
\definecolor{White}{gray}{1}
\usepackage{microtype}

\usepackage[ruled]{algorithm2e} 

\SetAlFnt{\small}
\SetAlCapFnt{\small}
\SetAlCapNameFnt{\small}
\SetAlCapHSkip{0pt}

\acmJournal{TOG}

\begin{document}
\title{Interactive and Robust Mesh Booleans}

\author{Gianmarco Cherchi}
\affiliation{%
	\institution{University of Cagliari}
	\country{Italy}
}

\author{Fabio Pellacini}
\email{pellacini@di.uniroma1.it}
\affiliation{%
	\institution{Sapienza University of Rome}
	\country{Italy}
}

\author{Marco Attene}
\email{marco.attene@ge.imati.cnr.it}
\affiliation{%
	\institution{CNR IMATI}
	\country{Italy}
}

\author{Marco Livesu}
\email{marco.livesu@gmail.com}
\affiliation{%
	\institution{CNR IMATI}
	\country{Italy}
}

\renewcommand\shortauthors{Cherchi, G. et al}

\newcommand{\giammi} [1] {\textbf{\color{orange} [Giammi]: #1}}
\newcommand{\cino} [1] {\textbf{\color{blue} [Cino]: #1}}
\newcommand{\fabio} [1] {\textbf{\color{green} [Fabio]: #1}}
\newcommand{\jaiko} [1] {\textbf{\color{magenta} [Jaiko]: #1}}

\begin{teaserfigure}
\centering
\includegraphics[width=\linewidth]{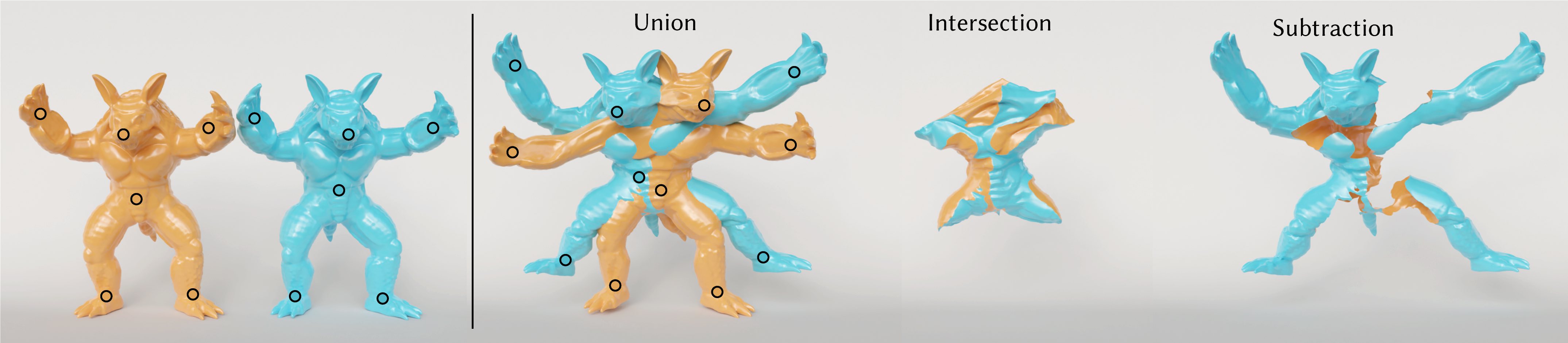}
\caption{We allow users to perform robust Boolean operations in real time on non trivial meshes containing thousands of triangles. In this example, the user first selects an arbitrary number of deformation handles (left). During the interactive session the handles can be freely moved in space, and the system applies both As-Rigid-As-Possible deformation~\cite{sorkine2007rigid} and our robust Booleans in real time. The two meshes of the armadillo contain 50K triangles each.}
\label{fig:arap_demo}
\end{teaserfigure}

\begin{abstract}
Boolean operations are among the most used paradigms to create and edit digital shapes. Despite being conceptually simple, the computation of mesh Booleans is notoriously challenging. Main issues come from numerical approximations that make the detection and processing of intersection points inconsistent and unreliable, exposing implementations based on floating point arithmetic to many kinds of degeneracy and failure. Numerical methods based on rational numbers or exact geometric predicates have the needed robustness guarantees, that are achieved at the cost of increased computation times that, as of today, has always restricted the use of robust mesh Booleans to offline applications.
We introduce the first algorithm for Boolean operations with robustness guarantees that is capable of operating at interactive frame rates on meshes with up to 200K triangles.
We evaluate our tool thoroughly, considering not only interactive applications but also batch processing of large collections of meshes, processing of huge meshes containing millions of elements and variadic Booleans of hundreds of shapes altogether. In all these experiments, we consistently outperform prior art by at least one order of magnitude.

\end{abstract}

%
%


%
%


\maketitle

\section{Introduction}
\label{sec:intro}

\begin{figure*}
	\centering
	\includegraphics[width=\linewidth]{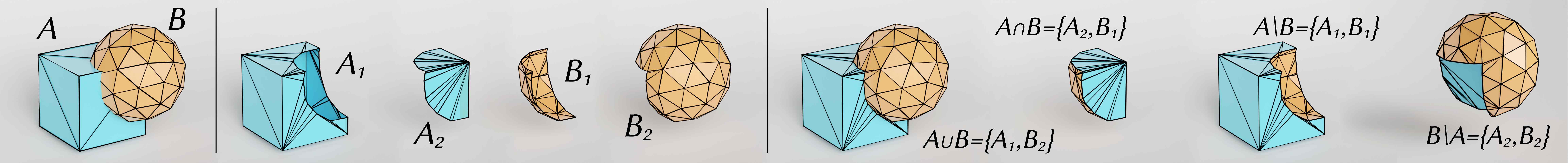}
	\caption{Computing a Boolean between two shapes amounts to: (i) resolve mesh intersections, creating a set of conforming surface patches; (ii) merge the patches to form the  output depending on the Boolean operator of choice.}
	\label{fig:pipeline}
\end{figure*}

\begin{figure}
	\centering
	\includegraphics[width=\columnwidth]{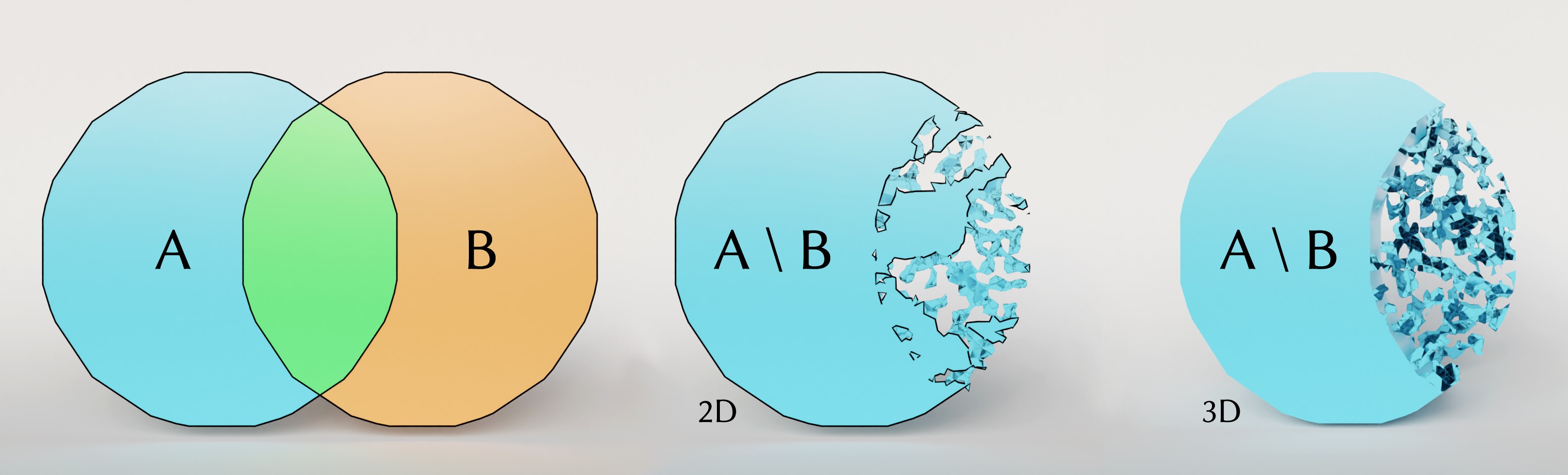}
	\caption{Two failure examples of Cork~\cite{cork}, one of the most robust floating point implementations of mesh Booleans. 
		The Euler characteristic of the 2D version (middle) of $A \setminus B$ is -50. The one of the 3D version (right) is 16. The Boolean with thickened disks was also tested on the the experimental tool available in Geogram/Graphite~\cite{graphite} -- which is restricted to solid objects -- leading the program to a crash.}
	\label{fig:cork_fail}
\end{figure}

Combining 3D meshes through Boolean operations is a fundamental functionality to define complex shapes via Constructive Solid Geometry (CSG). Despite being intuitive to the user, mesh Booleans are complex to implement correctly since small numerical errors may lead to unpredictable topological errors.
Interactive modeling software is mostly based on non-robust methods to achieve interactivity during modeling, but this leads to significant workflow problems for users, as anecdotally shown by the galleries of failure cases that populate user forums. 
Professional software such as Autodesk Maya
and Blender
openly report instability issues in presence of challenging configurations such as coplanarity~\cite{mayafail,blenderfail}.\\

Academic research has been studying the problem for a long period, developing robust algorithms and sometimes providing free-to-use implementations, such as~\cite{libigl}. To this day though, robust methods are still too slow to secure interactive frame rates relegating their use to offline modeling applications, e.g. for engineering and fabrication~\cite{nuvoli2019quadmixer,ureta2016interactive,garg2016computational,jacobson2017generalized,muntoni2018axis,alderighi2018metamolds,alderighi2019volume,dai2018support,fanni2018fabrication,yao2017interactive,attene2018exact}.\\
\looseness=-1


In many existing methods, the calculation of a mesh Boolean is framed as a two step process. In the first step, conflicts between mesh elements are resolved, splitting triangles in order to incorporate intersection lines in the connectivity. In the second step, each mesh element is deemed as being inside or outside each input object. The result of a Boolean is eventually computed as a subset of the mesh elements generated in the first step, filtered according to the inside/outside labeling computed in the second step. For example, the union $A \cup B$ of two triangle meshes $A$ and $B$  is the set of triangles in $A$ that are outside $B$ plus the triangles in $B$ that are outside $A$ (Figure~\ref{fig:pipeline}).\\


The major difficulty in implementing a Boolean pipeline comes from the use of finite precision arithmetic, which does not allow to exactly represent and test intersection points. In practice, this translates into a variety of artifacts that span from the definition of incomplete intersection lines to topologically inconsistent partitions that make the inside/outside relations ill-defined. A typical failure is shown in Figure~\ref{fig:cork_fail}, where the subtraction between two largely overlapping discs has significant issues.
Robust methods for inside/outside partitioning exist, but they are either computationally inefficient~\cite{jacobson2013robust} or are efficient at query time but are approximate and require initialization~\cite{barill2018fast}. The alternative is to eschew the use of floating point arithmetic and represent point coordinates with rational numbers~\cite{zhou2016mesh} or implicitly~\cite{diazzi2021convex}. These techniques ensure a topologically correct result at the cost of 1 to 2 orders of magnitude slow down w.r.t. floating point arithmetic, and are therefore not suitable for interactive use either.\\



In this paper we bring, for the first time, the robustness of exact methods into the world of interactive applications for general meshes, by improving the efficiency of exact algorithms by at least one order of magnitude compared to the state of the art, and without any pre-processing. As demonstrated in Section~\ref{sec:results}, this makes robust mesh Booleans available when interactively editing meshes with up to 200K triangles on commodity laptops. Besides interactive applications, our algorithm performs significantly better than the state of the art in many practically relevant offline applications, spanning from batch processing of large collections of data, Booleans between production-level high resolution meshes with millions of triangles, and variadic Booleans involving hundreds of input shapes.\\

Our improvements are made possible by significant contributions to both steps of the Boolean pipeline. For the first part, our method is based on a derivation of the mesh arrangements described in~\cite{CLSA20_arrangement}, which we improved as detailed in Section~\ref{sec:arrangements}, obtaining an average speedup of more than $5\times$. 
For the second part, we exploit the guaranteed topological correctness of the arrangement, coupling it with a robust ray casting approach that allows to reliably compute the inside/outside labels by throwing a single ray per patch. Taking inspiration from robust predicates~\cite{levy2016robustness,attene2020indirect,richard1997adaptive}, we formulate the inside/outside tests as a cascaded sequence of ray casting methods, sorted from faster but non-robust to slower but robust. We eventually resort to fully exact, thus fully robust, ray casting only when strictly necessary. As detailed in Section~\ref{sec:raytracing} this solution is up to $100\times$ faster than existing approaches based on topological flooding~\cite{attene2014direct} or patch graph processing~\cite{zhou2016mesh} used by previous exact methods, and also scales optimally to big meshes composed of millions of triangles and variadic Booleans involving hundreds of meshes (Section~\ref{sec:results}).\\

All in all, we believe this work considerably improves upon the state of the art in terms of numerical speedup and, perhaps more importantly, it brings robustness to interactive applications. Our experiments begin to explore ideas on how robust Booleans can be used in real-time, though this is just scratching the surface of what future interactive applications may be able to do. To support these experiments together with future research, we make our prototype implementation available as open source at \textit{<anonymous>}.



\section{Start of the art}
\label{sec:related}

Boolean compositions can be displayed without computing an explicit 3D model of the result. This is sufficient for visualizing the result, and for a few other specific applications~\cite{ray_traced,zanni2018hcsg,chen2022real}. In general though, modeling systems require an explicit representation of the Boolean composition and, depending on the target application, its calculation may need to be robust and exact. Existing algorithms can be classified based on the numerical model employed, e.g. floating point vs exact arithmetic, the geometric approach, e.g. surface vs volume-based, or the type of result they produce, e.g. exact vs approximated.

\subsection{Numerical models}
The easiest approach to implement Boolean operations is to rely on floating point arithmetic. Due to the hardware support, floating point numbers are by far the fastest approach and are indeed used widely in both academic~\cite{cork,graphite} and professional implementations, such as AutoDesk Maya and Blender. As mentioned in the introduction, round-off errors make this approach quite \emph{fragile}, which in turn may lead to significant topological errors, shown for example in Figure~\ref{fig:cork_fail}. On the other extreme, unconditional robustness may be obtained by replacing floating point numbers with exact number types, for example with rational numbers ~\cite{schifko2010industrial}. The use of exact arithmetic leads to unacceptable slowdown, by one or more orders of magnitude compared to floating point.

For some geometry processing tasks, robustness can be obtained by just guaranteeing that the program flow is exact independently of round-off errors. This is done by evaluating \emph{geometric predicates} exactly and quickly through arithmetic filtering~\cite{richard1997adaptive,levy2016robustness}. A typical predicate calculates the sign of a polynomial and, as long as the sign is correct, the program flow is guaranteed to be consistent. Arithmetic filtering~\cite{Devillers1998} makes it possible to evaluate a polynomial using floating point arithmetic but, along with it, an upper bound for the rounding error is computed. If the magnitude of the evaluated expression is larger than the error bound, its sign is guaranteed correct. If not, the filter \emph{fails}, and the predicate is re-evaluated using arbitrary precision. If the failure rate is low enough, absolute precision rarely comes into play and the slowdown is acceptable~\cite{magalhaes2017fast}, in particular when parallel architectures are employed \cite{de2020efficient}.

Arithmetic filtering can only be employed if the predicate input is guaranteed correct. It is therefore used only for tasks where the input comes from the ground truth, such as in mesh generation, where the predicate works directly on the coordinates of the input points~\cite{hang2015tetgen,shewchuk1996triangle}. For geometric tasks that rely on \emph{intermediate constructions}, the state of the art solution is lazy exact evaluation~\cite{PION2011}, which is still too slow for interactive applications. Luckily, for the case of mesh Booleans the only intermediate constructions are the intersection points. As recently shown in~\cite{attene2020indirect}, such points can be implicitly represented as the intersection of input primitives and the point's expression can be composed with the predicate's expression, enabling the use of arithmetic filtering. Our method exploits this latter technique to perform fast and exact geometric queries involving any combination of input and intersection points.

\subsection{Surface and volume-based methods}
Methods that implement one or both of the steps of the Boolean pipeline may work either with an explicit volumetric mesh or with a surface mesh enclosing the volumes of interest. In the general case, explicit volumetric representations make the algorithm easier (e.g. for in/out labeling) but are less efficient due to the increased dimensionality. Surface based methods are a bit more convoluted but in general more performing. Our method belongs to this latter category.

\paragraph{Volume-based}
Pioneering works supporting Boolean operations were mainly based on simple volumetric primitives (e.g., spheres, cylinders, half-spaces) or on implicit representations for more generic inputs~\cite{pasko1999hyperfun,wyvill1999extending}. Besides some notable exception~\cite{sellan2021swept}, the growing shape complexity in modern geometric modeling has gradually pushed the use of explicit mesh-based representations. Based on CGAL's exact kernel, in~\cite{hu2018tetwild} input meshes are used to partition the space into conforming volumetric cells, obtaining a mesh arrangement. Its faster version~\cite{hu2020fast} uses floating point calculations to create a conforming tetrahedral mesh, though with no formal guarantees. The creation of a volumetric mesh which conforms to the input was also used in~\cite{diazzi2021convex}, where exact arithmetic was replaced by indirect predicates and Boolean operations could be extracted at a much higher speed while maintaining the correctness guarantees.
The problem of partitioning the space based on input facets was also tackled in~\cite{paoluzzi2019finite,paoluzzi2020topological} based on floating point computation, using the language of geometric and algebraic topology. In~\cite{tao2019mandoline}, the fragility of floating point computation was reduced by using axis-aligned planes to cut the input triangles and define the volumetric cells, but being purely based on floats this method does not provide guarantees of correctness.

\paragraph{Surface-based}
Among surface-based methods, exact constructions are used in~\cite{schifko2010industrial} while walking on the input surfaces and splitting triangles when intersections are encountered. Since using exact constructions is expensive, \cite{attene2014direct} proposes a hybrid approach that is still based on walking on the outer surface, but uses floating point constructions whenever the rounding is harmless, while still switching to exact arithmetic when necessary. In~\cite{xu2013fastbools}, the topology of the resulting surface is guaranteed correct thanks to a clever use of orientation predicates with no need of exact constructions. 
Instead of walking on the surface, \cite{mei2013simple} uses a temporary octree to quickly find the candidate pairs of intersecting triangles. \cite{zhou2016mesh} exploit CGAL's exact kernel to partition the space into cells, each labeled with winding numbers w.r.t. the input. Boolean operations are performed by simply selecting a subset of the cells according to their labels. The algebraic composition of intersection points and predicates introduced in~\cite{attene2020indirect} was exploited in~\cite{CLSA20_arrangement} to quickly transform an arbitrarily self-intersecting soup of triangles into a well-formed simplicial complex. 
Despite being the fastest arrangement algorithm available, the original implementation of~\cite{CLSA20_arrangement} is still not sufficiently fast for interactive applications. The first step of our Boolean pipeline is heavily based on~\cite{CLSA20_arrangement}, but we substituted some of its modules and redesigned the data structures and algorithm flow paths to make it more amenable to parallel execution (Section~\ref{sec:arrangements}), obtaining an average speedup of $5\times$ (Section~\ref{sec:results}).

\subsection{Exact and approximated methods}
When applications tolerate an approximation, input vertex coordinates can be converted to triangle plane coefficients that can be easily combined to implement Boolean operations~\cite{bernstein2009fast}. Since the conversion is not exact, the result typically needs to be repaired. Other noticeable options to produce approximated Booleans are~\cite{barill2018fast} and~\cite{hu2020fast} where robustness and speed are combined. In~\cite{pavic2010hybrid}, regions close to the intersection lines are finely remeshed instead of being exactly cut.

In some cases, however, approximations cannot be tolerated. As an example, consider an interactive modeling system where a designer can perform numerous, subsequent and unpredictable operations: approximation would quickly accumulate and lead to poor results.
In these cases, exact results can be obtained based on slow arbitrary precision~\cite{schifko2010industrial,zhou2016mesh}, or by cleverly considering the floating point rounding.
In~\cite{campen2010exact} input edges are split so that they become short enough to guarantee that the coordinate-plane conversion in~\cite{bernstein2009fast} is lossless and can then be used to produce Boolean composition and other interesting modeling operations~\cite{campen2010polygonal} with no repairing. The problem with these methods is that the plane-based result becomes exceptionally complex when operations are cascaded. In an attempt to reduce this effect, in~\cite{sheng2018accelerated} both the coordinates and the planes are used to reduce the need for conversions. Alternatively, an effective approach to simplify intermediate plane-based representations has been recently reported in~\cite{nehring2021fast}.
In a recent trend of works, indirect predicates are used as a replacement for intermediate constructions when calculating mesh arrangements~\cite{CLSA20_arrangement} or polyhedral space subdivisions~\cite{diazzi2021convex}. When used to calculate mesh Booleans, this approach guarantees robustness without sacrificing speed. Nonetheless, if the explicit volume is not necessary for the final application, its calculation represents a useless overhead.

\paragraph{Snap rounding}
Though exact methods are clearly useful, they have a common problem when the result needs to be saved to a file or passed to other, non exact, algorithms. In these cases exact or implicit coordinates must be converted to inexact floating point values and the necessary approximation may invalidate the model by introducing degenerate or intersecting elements. A provably correct and efficient solution to this problem is still elusive and existing algorithms are either impractical~\cite{devillers_et_al18} or do not guarantee to produce a correct result in all the cases~\cite{milenkovic2019geometric}. However, existing heuristics proved to fail only in an extremely small percentage of practical cases~\cite{zhou2016mesh}. It is worth noticing that meshes created with exact methods endow topological properties that cannot be obtained by approximate methods. For example, the result of a Boolean operation between two watertight manifold meshes that do not touch tangentially is guaranteed to be manifold watertight. Properties of this kind may be relevant also for downstream applications, regardless of the geometric degeneracies that snap rounding may introduce in the output.

\subsection{In/out classification}
\label{sec:inout}
A common problem in most mesh Boolean algorithms is determining whether elements are part of the result or not. In surface-based methods, these elements are triangles and one needs to know whether they bound the result or not. In volume-based methods, elements are cells that might or might not be part of the (volumetric) result. A widely used approach for surface-based methods is to \emph{walk} on the surface and track the portions that belong to the result based on geometric reasoning~\cite{schifko2010industrial,attene2014direct}. Exact methods guarantee that the input arrangement is well-formed, hence cells in a volumetric decomposition can be easily classified by starting from the \emph{infinite} external cell and possibly switching from exterior to interior (and vice versa) when portions of the input surface are crossed. These approaches are clearly incompatible with naive floating point implementations, because of the lack of topological guarantees. When the input has surface holes or self-intersects in an ambiguous way, the concept of generalized winding number proved to be effective~\cite{jacobson2013robust}, though in some cases algorithms based on graph cuts provide better solutions~\cite{diazzi2021convex}. Since a naive computation of winding numbers is too slow in practice, a faster though approximate algorithm was proposed in~\cite{barill2018fast}. These methods are in general slower than topological walking, but constitute an unavoidable cost to pay for inexact methods based on naive floating points. The inside/outside classification system we use in this paper is based on exact ray casting (Section~\ref{sec:raytracing}) and proved to be much faster than both types of approaches (Section~\ref{sec:results}), also exhibiting a much better scalability on complex variadic Booleans containing hundreds of input shapes (Section~\ref{sec:variadic}).



\begin{figure*}
	\includegraphics[width=\linewidth]{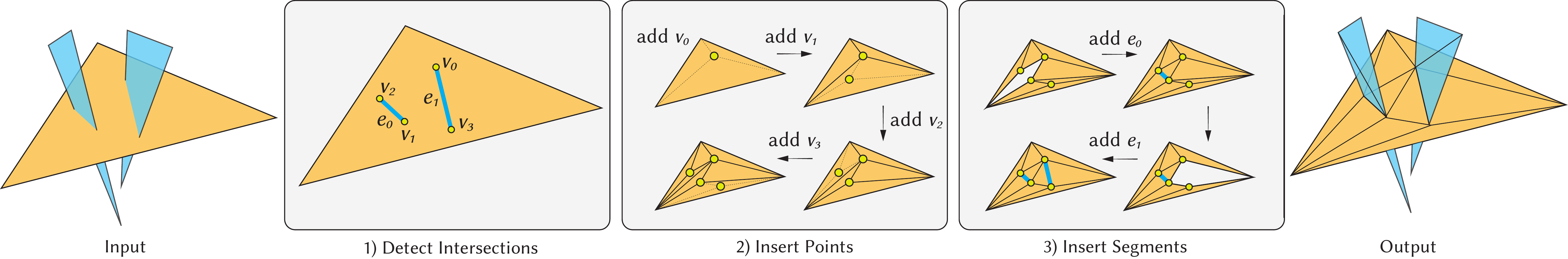}
	\vspace{-0.5cm}
	\caption{First step of the Boolean pipeline. We start from an generic triangle soup (left) and detect intersection points and lines (1). We then split triangles to incorporate new points first (2), and then segments (3). The output is a well formed simplicial complex (right).}
	\label{fig:arrangement}
\end{figure*}

\section{Overview}
\label{sec:overview}
Our method takes as input a set of input meshes $M_1,M_2,\dots,M_n$, and a Boolean operator, namely union, intersection, subtraction. Input meshes are always assumed to unambiguously enclose a volume, that is, they are manifold, watertight and with no self-intersections. The output is a mesh $B$ that contains the result of applying the Boolean operator to the input meshes. By definition of Boolean, $B$ is a sub-volume of the input, hence it is bounded by portions of input triangles. Mapping the result of a Boolean to the input shapes is useful in many applications, therefore for each output triangle we propagate information on its origin. Note that an output triangle can belong to \emph{many} input meshes, for example in the case of meshes that overlap at a coplanar region.\\
As anticipated in Section~\ref{sec:intro}, the Boolean algorithm may be regarded as a two-steps pipeline. In the first step, intersecting mesh elements are split and intersection lines are incorporated in the elements connectivity. With slight abuse of notation we refer to this step as \emph{mesh arrangement}~\cite{CLSA20_arrangement}. Throughout this paper, an arrangement is a set of surface patches bounded by intersection lines. When exact methods are used, the arrangement is guaranteed to be a well formed simplicial complex and surface patches are bounded by closed loops of non-manifold edges, namely the intersection lines. We take advantage of this property in the second phase of our algorithm, that takes the arrangement as input and processes each of its patches to determine whether they are positioned inside or outside with respect to each of the input meshes $M_1,M_2,\dots,M_n$. The output of the algorithm is eventually obtained by filtering the patches in the arrangement according to this information. For example, the union of two triangle meshes $M_1$ and $M_2$ ($M_1 \cup M_2$) is the set of patches of $M_1$ that are outside $M_2$ plus the patches of $M_2$ that are outside $M_1$. In the following two sections, we detail our technical contributions to each step of the pipeline. Visual examples for all Boolean operators we support are shown in Figure~\ref{fig:pipeline}.\\



\section{Mesh Arrangement} 
\label{sec:arrangements}

From the perspective of the arrangement algorithm, the input meshes $M_1,M_2,\dots,M_n$ can be seen as a soup of possibly intersecting triangles. We therefore flatten all input triangles into a single array, associating to each triangle a tag that maps it to the input mesh it belongs to.

At the highest level, all existing arrangement algorithms operate in a similar fashion, detecting intersections between triangles first, and eventually proceeding with mesh refinement, inserting intersection points first, and then segments~\cite{zhou2016mesh,attene2014direct,CLSA20_arrangement} (Figure~\ref{fig:arrangement}). Differences between the various methods are in the fine technical details, such as how intersection points are represented and processed, or how segment insertion is performed. These choices are fundamental to ensure that the algorithm is fast, memory efficient, amenable to parallelization and able to scale well on large datasets. We based our implementation of the arrangement step on the method described in~\cite{CLSA20_arrangement}. Even though this is the fastest existing method in its class, it is not fast enough for interactive use on our target mesh size. We therefore introduced a few important improvements to the original pipeline, enhancing the detection of intersections and the insertion of intersection lines (steps 1 and 3 in Figure~\ref{fig:arrangement}). Overall, we obtained a speed up factor of $3-8\times$ w.r.t. the original algorithm of~\cite{CLSA20_arrangement}. In the remainder of this section we detail all the major improvements that we introduced. For a broader discussion of the whole algorithm we refer the reader to the original article.

\paragraph{Cached Predicates.} To obtain unconditional numerical robustness all operations involving the detection of intersections, point in triangle location for vertex insertion and re-triangulation for segment insertion (steps 1,2,3 in Figure~\ref{fig:arrangement}) must be based on exact orientation predicates~\cite{richard1997adaptive}, which therefore constitute a computational bottleneck for the arrangement algorithm. The most frequent operation is the so called \verb+orient3D+, which locates a point in space with respect to a given plane. Given a point $p$ and a plane passing through points $a,b,c$, the orientation amounts to computing the sign of the determinant
$$
orient3D(a, b, c, p) =
\left\vert
\begin{array}{cccc}
a_x & a_y & a_z & 1\\
b_x & b_y & b_z & 1\\
c_x & c_y & c_z & 1\\
p_x & p_y & p_z & 1\\
\end{array}
\right\vert
$$
In its classical form, this determinant is evaluated from scratch at each predicate call. We observe that in the arrangement algorithm planes are known a priori (they are the supporting planes of the input triangles) and are tested multiple times against several different points. It is therefore convenient to compute the plane based portion of the determinant once and to use it each time the same plane is tested against a new point. Starting from this intuition, we rewrite the $4 \times 4$ determinant above as
\begin{align*}
orient3D(a, b, c, p) = &
-p_x \left \vert \begin{array}{ccc}
a_y & a_z & 1\\
b_y & b_z & 1\\
c_y & c_z & 1\\
\end{array} \right \vert
+ p_y \left \vert \begin{array}{ccc}
a_x & a_z & 1\\
b_x & b_z & 1\\
c_x & c_z & 1\\
\end{array} \right \vert\\
& - p_z \left \vert \begin{array}{ccc}
a_x & a_y & 1\\
b_x & b_y & 1\\
c_x & c_y & 1\\
\end{array} \right \vert
+ \left \vert \begin{array}{ccc}
a_x & a_y & a_z\\
b_x & b_y & b_z\\
c_x & c_y & c_z\\
\end{array} \right \vert
\end{align*}
thus obtaining a perfect separation between plane coefficients and point coordinates. We exploit this latter equation to cache, for each input triangle, the four $3 \times 3$ determinants, thus reducing each call to \verb+orient3D+ to a simple scalar product in 4D. 
Similar caching techniques were recently exploited for the tetrahedralizations of huge point clouds composed of billions of points~\cite{marot2019one} and are also used in Boolean pipelines based on plane representations~\cite{nehring2021fast}.

\paragraph{Segment Insertion.} To make sure that intersection lines are correctly incorporated in the output mesh, not only intersection vertices but also intersection segments must be inserted (step 3 in Figure~\ref{fig:arrangement}). Inserting a segment amounts to eliminating, from the current tessellation, all triangles that conflict with it, and then re-triangulate the so-generated polygonal pocket, while making sure that the wanted segment is part of the new tessellation. This is a classical yet widely studied problem in computational geometry~\cite{shewchuk2015fast}. In~\cite{CLSA20_arrangement} segment insertion was performed using the earcut algorithm, which in its best implementation achieves $O(n^2)$ worst case complexity~\cite{eberly2008triangulation}, with $n$ being the number of polygon segments. We substituted earcut with a method recently introduced in~\cite{livesu2021deterministic}, which ensures optimal deterministic $O(n)$ complexity in all cases and is two orders of magnitude faster than the previously best performing existing method~\cite{shewchuk2015fast}.

\paragraph{Low-level Implementation.}
To further improve execution speed, we redesigned the underlying data structure to increase the possibilities for parallel execution. Overall, we observe that the most costly operations of the arrangement algorithm are the removal of duplicate and degenerate elements, the construction of adjacencies, the intersection computations and the final triangulation. The main costs of these operations can be reduced by executing their core components in a data-parallel manner, giving us high speed up while maintaining the same algorithm pipeline. We also optimize octree construction by using nested parallel constructs. Parallel constructs are executed by a work-stealing scheduler that ensures good balanced workloads. Note that we perform degeneracy removal and rebuilding of adjacencies each frame to ensure robustness while allowing any modeling operation to be performed by users. We considered additional parallelization opportunities in patch construction and final mesh extraction. In these case though, parallelism is harder to extract since it requires fine-grained locking. In fact, we tested these approaches but without gaining speed up, suggesting that a meshlet-based approach is likely needed to extract further parallelism from the pipeline~\cite{mahmoud2021rxmesh}.

The final improvement we implemented was the use of specialized data structures to improve cache coherency, reduce memory fragmentation due to deletions, and reduce the overall pressure on the system memory allocator. Regardless of the data structure used to store the mesh, and the size of the mesh, many small memory operations, including deletions, are required to update the data structure, and these updates have significant performance implications. In the arrangement step, the most expensive low-level data structures are sets, dictionaries and sparse graphs stored as adjacency lists. We optimize these data structures with three techniques. First, we use hash tables based on the swiss table design for sets and dictionaries, to both save memory and improve cache coherency. Second, we use arena allocators to reduce the pressure on the memory allocator and reduce overall fragmentation. Third, we use dynamic arrays with small-array optimization for adjacency lists. Overall these techniques provide a relevant speedup throughout the pipeline.




\section{Inside/outside classification}
\label{sec:raytracing}
\begin{figure}
	\centering
	\includegraphics[width=\linewidth]{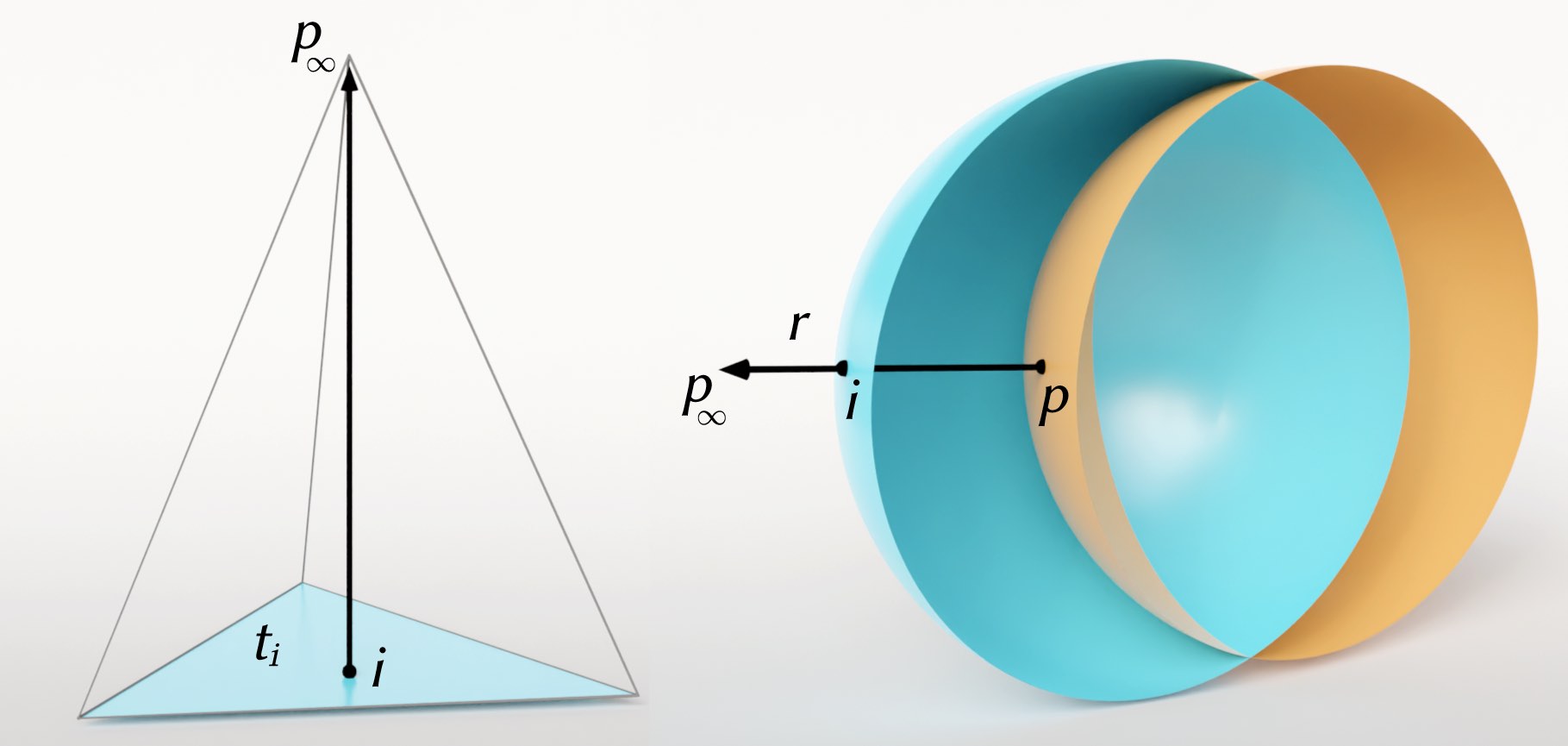}
	\caption{For each patch in the arrangement we robustly devise inside/outside relationships with input shapes with exact ray casting. We shoot a ray towards infinite and analyze its first intersection with all the input meshes. Given an intersection point $i$ and the triangle $t_i$ containing it, the ray traverses the mesh from inside to outside if the volume of the tetrahedron $(t_i,p_\infty)$ is negative, from outside to inside otherwise. This check can be performed exactly with arithmetic orientation predicates.}
	\label{fig:raycasting}
\end{figure}

In this second phase, we consider the arrangement computed at the previous step and determine, for each of its manifold surface patches, the relative position with respect to the input meshes $M_1,\dots,M_n$. Differently from the arrangement step, which is an amelioration of an existing technique, the computation of the inside/outside classification is entirely different from the topological approaches used in prior art~\cite{attene2014direct,zhou2016mesh}. Our key insight is that the inside/outside relationship between a patch and an input mesh $M$ can be determined by casting a ray from any patch point along an arbitrary direction and then analyzing its intersection with $M$ (Figure~\ref{fig:raycasting}, right). An important aspect of such an approach is that the algorithm scales with the number of patches in the arrangement and not with the number of triangles in the mesh. Since the former is typically orders of magnitude lower than the latter and the cost of casting a single ray is almost negligible, the algorithm becomes remarkably fast. In our tests, we achieved up to $100\times$ speedup compared to prior art, while also showing better scalability on variadic Booleans involving numerous input shapes (Section~\ref{sec:results}).\\

Algorithm~\ref{alg:boolean_pipeline} summarizes the main steps of our ray casting approach. For each input patch $P$, we construct a ray $r$ that emanates from it and points towards a point at infinity $p_\infty$ that is guaranteed to stay outside of all input meshes. We then test intersections between $r$ and each input mesh $M$. If at least an intersection occurs, we select the \textit{first} of them, that is, the one that is closest to the point of $P$ from where $r$ emanates.
To decide whether $r$ traverses $M$ from the inside to the outside or vice-versa, we compare the ray direction with the outfacing local surface normal. As shown in Figure~\ref{fig:raycasting}~(left), such a check translates into the evaluation of the signed volume of a tetrahedron, which can be computed exactly using orientation predicates. 


\begin{algorithm}[tb] 
	\scriptsize
	\textbf{Input:} input meshes $M_1,\dots, M_n$ and their arrangement patches $P_1,\dots,P_m$\\
    \textbf{Output:} relative position of patches $P_1,\dots,P_m$ w.r.t. the input meshes $M_1,\dots, M_n$\\
	\hrulefill \\
	\SetAlgoLined
	\vspace{0.1em}
	\For{each patch $P$}
	{
		Initialize $P$ as being outside of $M_1,\dots, M_n$\;
        Define a ray $r$ starting at point $p \in P$ towards point at infinite $p_\infty$; \hfill (Sec.~\ref{sec:ray_definition})\\
		\For{each input mesh $M$}
		{		
		    compute and sort intersections between $r$ and $M$; 	\hfill (Sec.~\ref{sec:ray_intersection})\\
			\If{$r$ and $M$ intersect} 
			{
				find intersecting triangle $t \in M$; \hfill (Sec.~\ref{sec:classification})\\
				compute volume of tetrahedron $(t,p_\infty)$; \hfill (Fig.~\ref{fig:raycasting})\\
				\If{volume is negative} 
				{
					set $P$ as being inside $M$\;
				}            
			}
		}
	}
	\caption{Inside/outside classification}
	\label{alg:boolean_pipeline}
\end{algorithm}

This ray casting approach poses several technical challenges. First, it can only be applied to \emph{exact} arrangements computed with robust predicates or rational numbers, because alternative non-robust techniques cannot guarantee the absence of gaps or tiny topological channels connecting different patches. 
 Second, intersection detection must be \emph{exact} as well, because approximations in the computation may introduce artificial intersections or miss existing ones. Third, ambiguities that arise when the ray and the surface are tangent must be properly handled to ensure the correctness of the result. To make things even worse, we recall that intersection points inserted during the construction of the arrangement do not have known explicit coordinates, therefore computations must be fully compatible with implicit point representations. Note that this process is intrinsically unstable and demands absolute precision. A ray that misses an intersection because of a tiny geometric or topological imperfection may produce a wrong classification for an arbitrarily big patch. In the remainder of the section, we detail the main algorithmic steps, addressing all these aspects.

\begin{figure*}
	\centering
	\includegraphics[width=\linewidth]{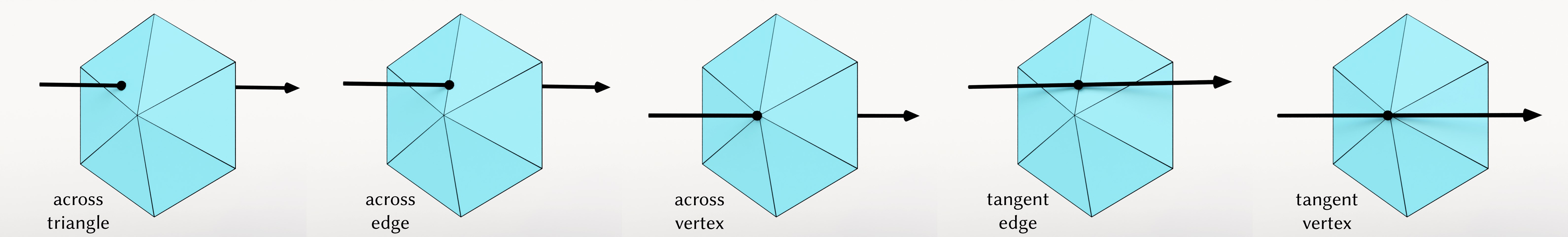}
	\caption{Five alternative cases of intersection between a ray and a triangle mesh. When the ray crosses the surface at a point that is inside a triangle (left) the test depicted in Figure~\ref{fig:raycasting} can reliably determine the inside/outside relation between the patch being tested and the surface crossed. When the ray hits a vertex or an edge the check becomes ambiguous. Our method always reconducts to the leftmost case via numerical perturbation of the ray. Note that the pathological cases depicted in the figure are not exhaustive. In fact, rays may also be tangent at a (coplanar) triangle.}
	\label{fig:inters_cases}
\end{figure*}




\subsection{Ray definition}
\label{sec:ray_definition}
Given a patch $P$, we need to define a ray that starts at a point $p \in P$ and passes through an infinite point $p_\infty$. If $p$ is known, the infinite point $p_\infty$ can be easily defined by translating $p$ along one of the major axes by a quantity that is bigger than the extent of the bounding box of the input scene along the same axis (this guarantees that $p_\infty$ is outside all input meshes). Moreover, having an axis-aligned ray considerably simplifies the next ray-triangle intersection analysis as it always allows to drop one coordinate and operate on a 2D plane instead of the 3D space. 

The main difficulty in this phase is the definition of the emanating point $p$. In the simplest case, the patch $P$ contains at least one input vertex with known ground-truth floating point coordinates in its interior, which can be used as a starting point for $r$. However, the arrangement may also contain patches that do not include input vertices or do not contain interior vertices at all. In all these cases, defining explicit floating point coordinates for $p$ may be intrinsically impossible as any rounding risks to detach the ray $r$ from the patch $P$ and possibly trigger errors in the classification. As mentioned in previous sections, numerical issues may be fully avoided by switching to costly rational numbers to represent point coordinates, at the cost of a major slowdown. Instead, inspired by the cascaded approaches used by filtered predicates~\cite{richard1997adaptive,levy2016robustness,attene2020indirect}, we define the ray $r$ by first attempting to find a satisfactory approximate floating point solution, while we resort to guaranteed exact rational numbers only as backup strategy. Our main idea is that if we can define a ray that starts from \emph{beneath} $P$ and is guaranteed to traverse the patch at some internal point $p \in P$, we can simply sort all the intersections we find, and perform the inside/outside classification by considering the first intersection that occurs \emph{after} $p$.

Our strategy is as follows: we pick a random triangle $t \in P$ and convert the coordinates of its implicit vertices into explicit floats, computing the approximate triangle barycenter $b_t$. To make sure that $b_t$ stays beneath triangle $t$ we push the point backwards along the same axis used to define $p_\infty$. If the snap rounding of the triangle vertices succeeds, the ray starting from $b_t$ and passing through $p_\infty$ intersects triangle $t$, hence $P$, giving us a ray with which to perform the in/out classification. Unfortunately, this intersection is not guaranteed to exist because the conversion to floating point coordinates may move the ray away. 
Thanks to the axis alignment we can efficiently and reliably test that the wanted intersection exists with a simple 2D point in triangle test, performed considering the orthogonal projection of both $t$ and $b_t$ along the ray direction. If the test fails we attempt to produce the same construction with another triangle of $P$, until we find a valid one. If no valid triangle can be found, we compute the exact barycenter with rational numbers and perform an exact ray casting. In practice, patches that do not contain input vertices are rare, and failures of our approximate strategy are even more so. In our large scale benchmark (Section~\ref{sec:benchmark}) we tested 3.8K Booleans, shooting more than 80K rays overall. Only the $2\%$ of these rays necessitated to perform snap rounding. In the $97\%$ of these cases we successfully defined a valid ray at the first triangle we tried. In $2\%$ of the cases we tested two triangles, and in $0.56\%$ of the cases we tested three triangles. In the worst case, we tested five triangles per patch, and we never hit the last step of our cascaded approach based on rational numbers.

\subsection{Intersection detection}
\label{sec:ray_intersection}
In a typical ray casting implementation, ray-triangle intersections are computed using the efficient M{\"o}ller–Trumbore algorithm~\shortcite{moller1997fast}. Unfortunately, our exactness requirements make it impossible to rely on such an algorithm, which involves floating point operations that accumulate error and it is also non stable in case of coplanarity. For the same reason, acceleration data structures that rely on non axis aligned planes or spatial hashing cannot be used, because ray intersection queries would require arithmetic operations that introduce unwanted approximation errors. In our implementation, we use a plain octree as acceleration structure and perform ray casting by testing intersections between the ray bounding box and each octant. Note that, for efficiency, this is the same acceleration structure used in the arrangement part to detect triangle intersections. Since both the octree and the ray are axis aligned we have two nice properties: the bounding box of the ray is tight (it's the ray itself), and the intersection with the octant reduces to a 2D check which involves only four comparisons between floats. For each leaf octant intersected by the ray we test all the triangles it contains. Once again, we exploit the axis aligned nature of our problem to recast the ray-triangle intersection as a point in triangle query in 2D, as previously described in Section~\ref{sec:ray_definition}. Thanks to this simplified formulation, our exact ray-triangle intersection routine becomes extremely fast. 

Since our classification is entirely based on the analysis of the first intersection between the ray and each input mesh, it is necessary to \emph{sort} intersection points. Firstly, we represent intersections implicitly, using the LPI (Line-Plane Intersection) points described in~\cite{attene2020indirect}. Then, we use the exact comparator introduced in~\cite{CLSA20_arrangement} to sort them from the closest to the ray emanating point to the furthest.

\subsection{Classification}
\label{sec:classification}
The first intersection between a ray $r$ and a mesh $M$ may take place in different ways. The simplest configuration is when $r$ crosses $M$ at a point that is interior to a triangle (Figure~\ref{fig:inters_cases}, left). 
In this case determining if the ray is passing from inside to outside or vice-versa consists in simply analyzing the triangle orientation, which is encoded in the local winding (triangle vertex order) and can be tested exactly as shown in Figure~\ref{fig:raycasting} (left). The cases when the ray intersects a mesh edge or a vertex are more difficult, because the ray may traverse them tangentially without crossing the surface, and because even if the surface is crossed, the choice of the triangle to test is ambiguous. Four of these cases are depicted in Figure~\ref{fig:inters_cases}, but there are also others (e.g. when a tangent ray is also coplanar to a triangle). While exact geometric tests could be performed to distinguish between all these cases, the computational overhead for detection and handling of pathological cases can be substituted with a simpler yet effective strategy.
Each time a ray does not intersect a patch triangle at an inner point, we perturb the coordinates of $p_\infty$ by $\epsilon$ (without moving its starting point) until the crossing happens at a point that is interior to a mesh triangle. Perturbation of point coordinates is performed using the next floating-point number representable starting from a given number (using \verb+std::nextafter+). The perturbed ray is then tested again for intersection, and the operation is repeated until a valid intersection point is found. Note that perturbed rays are no longer axis aligned, therefore we cannot reduce the ray-triangle intersection to a 2D problem, and we need to perform this test in 3D. The full test consists in checking the sign of three tetrahedra, formed considering the two ray endpoints and the endpoints of each triangle edge. If all signs are strictly positive, or negative, there is an intersection inside the triangle. The computational overhead of this check is minimal (three \verb+orient2D+ for the 2D case and three \verb+orient3D+ for the 3D case).

Intersections at mesh vertices and edges are extremely unlikely to happen in real shapes. In our large scale benchmark (Section~\ref{sec:benchmark}), we tested 3.8K Booleans, shooting more than 80K rays overall. In no case we performed numerical perturbation because all intersections occurred inside mesh triangles. We were only able to validate this code on an artificial example, obtained by performing a Boolean operation between a mesh and a copy of it translated along the $X$ axis. In this case all rays emanating from the patches of a mesh traverse the vertices of the other mesh. Indeed, a single perturbation was always sufficient to break ties and move the intersection point inside a triangle.

\section{Discussion}
\label{sec:results}

We implemented our Boolean pipeline in C++, using CinoLib~\cite{cinolib} data structures for exact ray casting and detection of intersections and Indirect Predicates~\cite{attene2020indirect} to robustly query intersection points. For maximum efficiency and parallel performance our code also relies on Google's Abseil fast containers and Intel's TBB.

We tested our algorithm thoroughly, considering interactive applications, batch processing of large collections of data, Booleans between huge meshes composed of millions of triangles, and variadic Booleans of hundreds of input meshes altogether.

Our baseline for comparative analysis is the method of~\cite{zhou2016mesh}, which is the latest released fully fledged exact Boolean pipeline. Since the original authors' implementation was released, the codebase underwent various improvements, also very recently. Unless specified differently, all numbers we report refer to the most recent implementation available in libigl~\cite{libigl}. We also compare to \cite{CLSA20_arrangement}, but this method covers only the first step of the Boolean pipeline. 
Whenever appropriate, we compare against a hybrid pipeline consisting of~\cite{CLSA20_arrangement} for the arrangement part and our ray casting (Section~\ref{sec:raytracing}) for inside/outside labeling.
Non-robust alternatives such as~\cite{cork,graphite} are not considered because they do not guarantee the topological correctness of the result and are prone to failures (Figure~\ref{fig:cork_fail}). 

As detailed in the remainder of this section, our Boolean algorithm proved to be superior than the state of the art by at least one order of magnitude in all experiments and is the only existing exact method capable of sustaining interactive frame rates for real-time applications.


\begin{figure*}
\centering
\includegraphics[width=\linewidth]{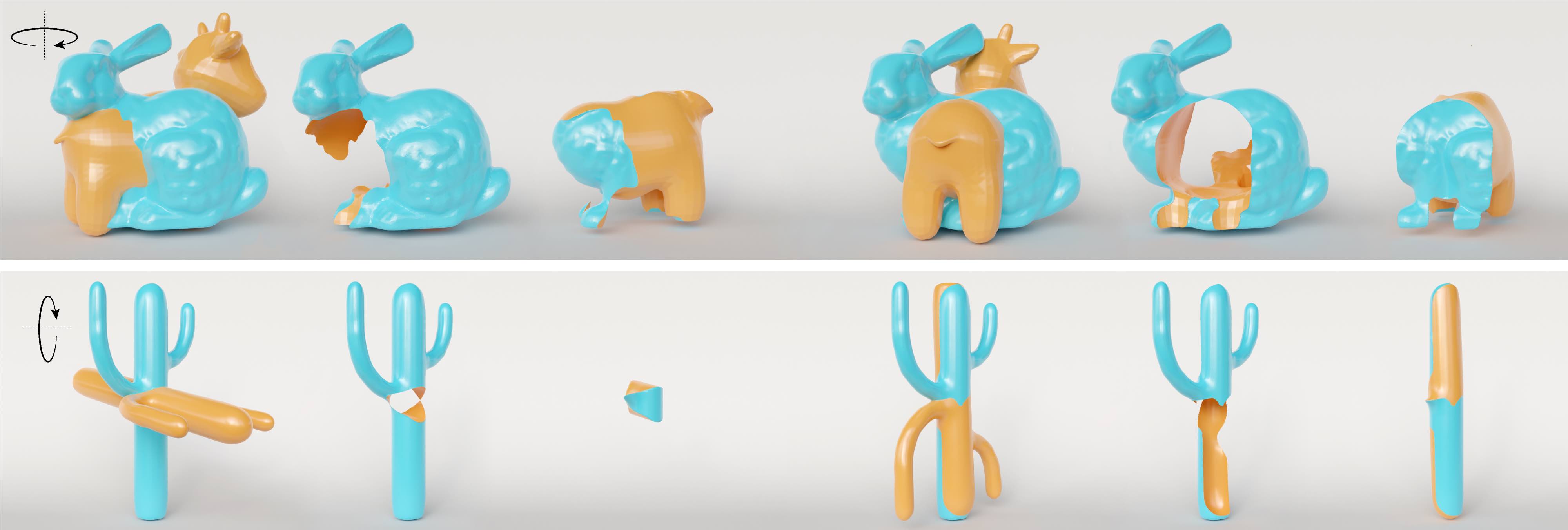}
\caption{Two examples of our interactive rotation demo: one mesh rotates on top of the other while the system executes a Boolean operator in real time. }
\label{fig:rot_demo}
\end{figure*}


\subsection{Interactive Applications}
\label{sec:interapp}
We considered both simple tasks where a scripted animation plays over time and fully dynamic tasks where all objects in the scene evolve over time in response of a user action. All interactive experiments were executed on a commodity laptop, a MacBook with M1 Pro with 8 performance cores and 32GB of RAM. Screen captures have been attached to the submission and are available to the reader to better judge the smoothness of the animation. Booleans are applied \emph{naively}, meaning that each frame is computed separately, without propagating cached data from one frame to the subsequent. We leave this improvement for future work to obtain additional speedups.

\paragraph{Rotation demo.} In this first test, two objects are rotated with respect to one another, while our algorithm computes the Boolean between the two, as shown in Figure~\ref{fig:rot_demo}. The user can interact with the system using the keyboard, selecting the type of Boolean operator between union, intersection and difference. 
This is the easiest interactive scenario: both objects are static up to a rigid movement of one of them. For maximum efficiency rendering and Booleans run in separate threads and are synchronized with double buffering.
We considered scenes of growing sizes, in the range from 25K to 200K triangles. Based on our experience the way shapes intersect to each other has a non negligible impact on running times. For example, Booleans become increasingly complex when the two shapes are almost perfectly aligned, and progressively simpler otherwise. To reduce any possible bias we always used the same two models, remeshed at various resolutions using Graphite~\shortcite{graphite}. Timings for each scene are reported in Table~\ref{tab:fps_rot}. As can be noticed, existing exact pipelines such as~\cite{zhou2016mesh} have computation times that are incompatible with interactive use already at the coarsest resolutions, running at roughly 1–2 fps for just 50K triangles. To give a comparative reference, we run at similar fps for 1M triangles. Using \cite{CLSA20_arrangement} for the first part of the pipeline yields running times that are compatible with interactive use for the coarsest resolutions, but the software is still not fast enough to scale on meshes containing more than 25-30K triangles. Our software runs interactively for scenes containing up to roughly 100-120K triangles, and starts to lag a bit for larger scene sizes. We point the reader to the attached screen captures to get a better sense of the smoothness of the animation.


\begin{table}
	\centering
	\resizebox{.85\columnwidth}{!}{
		\begin{tabular}{rccc}
			\toprule
			\textbf{Size} & \textbf{libigl ($t$)} & \textbf{FA($t$)}  & \textbf{Ours ($t$)}\\
			\midrule
			\rowcolor{Gray}
			25K & 0.22/0.78/0.39 &   0.03/0.12/0.07   &   0.01/0.04/0.02 \\
			50K & 0.36/1.58/0.50 &   0.12/0.20/0.13    &   0.02/0.06/0.04 \\
			\rowcolor{Gray}
			100K & 0.69/3.21/0.77 & 0.23/0.33/0.25    &   0.03/0.11/0.07 \\
			150K & 1.03/4.86/1.13 &  0.34/0.45/0.36   &   0.05/0.14/0.10 \\
			\rowcolor{Gray}
			200K & 1.37/6.48/1.50 & 0.46/0.62/0.50   &   0.06/0.19/0.14 \\
			\bottomrule
		\end{tabular}
	}
	\caption{Performances of the interactive rotation demo. For each scene we measure its size as the cumulative number of triangles and report minimum, maximum and average time per frame over a continuous run of 3 minutes. With libigl we denote the most recent implementation of~\cite{zhou2016mesh}. FA is a hybrid pipeline that uses the original implementation of the arrangement in~\cite{CLSA20_arrangement} with our in/out classification based on ray casting. All timings are in seconds.}
\label{tab:fps_rot}
\vspace{-0.5cm}
\end{table}

\begin{table}
	\centering
	\resizebox{.45\columnwidth}{!}{
		\begin{tabular}{rccc}
			\toprule
			\textbf{Size} & \textbf{min}  & \textbf{max} & \textbf{avg}\\
			\midrule
			\rowcolor{Gray}
			25K & 0.028 & 0.055 & 0.032 \\
			50K & 0.043 & 0.079 & 0.054\\
			\rowcolor{Gray}
			100K & 0.085 & 0.119 & 0.095\\
			150K & 0.133 & 0.164 & 0.140\\
			\rowcolor{Gray}
			200K & 0.188 & 0.215 & 0.195\\
			\bottomrule
		\end{tabular}
	}
	\caption{Performances of the interactive deformation demo. Timings (in seconds) report on the cumulative time to perform four ARAP iterations and our Boolean algorithm.}
	\label{tab:fps_arap}
	\vspace{-1.2cm}
\end{table}

\begin{figure*}
	\centering
	\includegraphics[width=\linewidth]{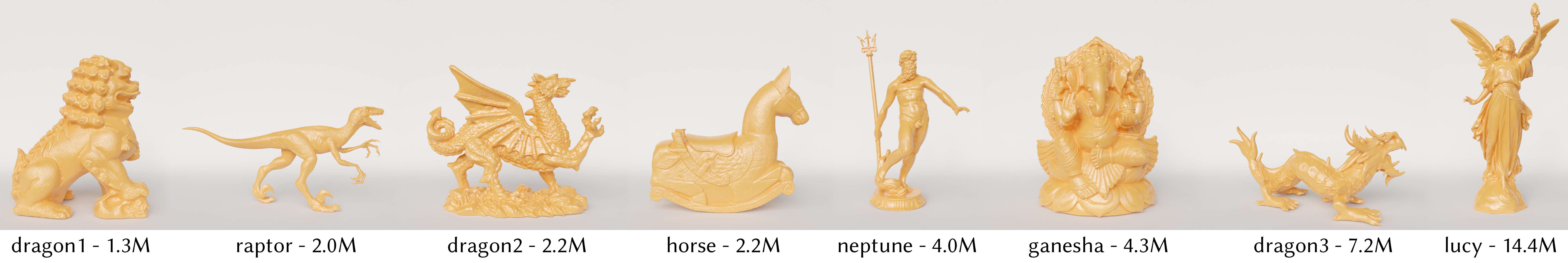}
	\caption{Dataset of big meshes we considered for the experiments in Table~\ref{tab:bigmeshes}.}
	\label{fig:bigmeshes}
\end{figure*}

\begin{table*}
	\centering
	\resizebox{.93\linewidth}{!}{%
		\begin{tabular}{llrcrrrcccrrrcrrrcc}
			\toprule
			\multicolumn{3}{c}{\textbf{Input}} & \phantom{aaa} &
			\multicolumn{3}{c}{\textbf{libigl ($t$)}} & \phantom{aaa} &
			\multicolumn{1}{c}{\textbf{FA ($t$)}} & \phantom{aaa} &
			\multicolumn{3}{c}{\textbf{Ours ($t$)}} & \phantom{aaa} &
			\multicolumn{3}{c}{\textbf{libigl ($\times$)}} & \phantom{aaa} &
			\multicolumn{1}{c}{\textbf{FA ($\times$)}} \\
			\cmidrule{1-3} \cmidrule{5-7} \cmidrule{9-9} \cmidrule{11-13} \cmidrule{15-17} \cmidrule{19-19}
			\textbf{obj 1}     & \textbf{obj 2}          & \textbf{size} && \textbf{arr}   & \textbf{bool}   & \textbf{tot}   && \textbf{arr}  && \textbf{arr}  & \textbf{bool} & \textbf{tot}  && \textbf{arr}   & \textbf{bool}  & \textbf{tot}   && \textbf{arr} \\
			\midrule
			dragon3 	& lucy      & 21.6M && 142.81 	& 46.86 	& 189.67 	&& 97.49 && 12.90 	& 6.61 	& 19.50 	&& 11.07$\times$ 	& 7.09$\times$ 	& 9.73$\times$ 	&& 7.56$\times$\\
			\rowcolor{Gray}
			lucy 	    & neptune 	& 18.4M && 131.33 	& 42.19 	& 173.52 	&& 79.87 && 12.00 	& 6.13 	& 18.13 	&& 10.95$\times$ 	& 6.88$\times$ 	& 9.57$\times$ 	&& 6.66$\times$\\
			ganesha 	& lucy      & 18.7M && 118.19 	& 427.02 	& 545.21 	&& 84.02 && 11.30 	& 6.04 	& 17.34 	&& 10.46$\times$ 	& 70.76$\times$ 	& 31.45$\times$ 	&& 7.43$\times$\\
			\rowcolor{Gray}
			lucy        & dragon2 	& 16.6M && 117.48 	& 412.19 	& 529.68 	&& 80.10 && 10.63 	& 5.38 	& 16.01 	&& 11.05$\times$ 	& 76.67$\times$ 	& 33.08$\times$ 	&& 7.53$\times$\\
			horse       & lucy      & 16.6M && 116.03 	& 35.83 	& 151.87 	&& 74.20 && 10.41 	& 5.27 	& 15.68 	&& 11.14$\times$ 	& 6.80$\times$ 	& 9.68$\times$ 	&& 7.12$\times$\\
			\rowcolor{Gray}
			lucy        & raptor 	& 16.4M && 110.12 	& 384.48 	& 494.60 	&& 68.10 && 10.01 	& 5.27 	& 15.28 	&& 11.00$\times$ 	& 72.94$\times$ 	& 32.37$\times$ 	&& 6.80$\times$\\
			dragon1 	& lucy      & 15.7M && 110.27 	& 30.12 	& 140.40 	&& 69.26 && 9.60 	& 4.90 	& 14.49 	&& 11.49$\times$ 	& \textcolor{red}{\textbf{6.15$\times$}} 	& 9.69$\times$ 	&& 7.22$\times$\\
			\rowcolor{Gray}
			dragon3 	& ganesha 	& 11.5M && 65.28 	& 242.44 	& 307.72 	&& 31.69 && 6.25 	& 3.09 	& 9.34 	&& 10.44$\times$ 	& 78.38$\times$ 	& 32.93$\times$ 	&& 5.07$\times$\\
			dragon3 	& neptune 	& 11.2M && 66.66 	& 20.78 	& 87.44 	&& 38.56 && 6.15 	& 2.98 	& 9.12 	&& 10.84$\times$ 	& 6.98$\times$ 	& 9.58$\times$ 	&& 6.27$\times$\\
			\rowcolor{Gray}
			dragon3 	& horse 	& 9.4M 	&& 56.25 	& 15.03 	& 71.28 	&& 26.62 && 5.19 	& 2.31 	& 7.49 	&& 10.85$\times$ 	& 6.51$\times$ 	& \textcolor{red}{\textbf{9.51$\times$}} 	&& 5.13$\times$\\
			dragon3 	& dragon2 	& 9.4M 	&& 54.84 	& 196.39 	& 251.23 	&& 30.46 && 5.00 	& 2.38 	& 7.38 	&& 10.98$\times$ 	& 82.38$\times$ 	& 34.04$\times$ 	&& 6.10$\times$\\
			\rowcolor{Gray}
			dragon3 	& raptor 	& 9.2M 	&& 50.31 	& 193.76 	& 244.07 	&& 21.82 && 4.70 	& 2.27 	& 6.97 	&& 10.71$\times$ 	& 85.28$\times$ 	& 35.02$\times$ 	&& 4.65$\times$\\
			ganesha 	& neptune 	& 8.3M 	&& 47.39 	& 174.23 	& 221.62 	&& 25.14 && 4.49 	& 2.27 	& 6.75 	&& 10.57$\times$ 	& 76.82$\times$ 	& 32.82$\times$ 	&& 5.61$\times$\\
			\rowcolor{Gray}
			dragon1 	& dragon3 	& 8.5M 	&& 51.05 	& 12.95 	& 64.00 	&& 27.82 && 4.67 	& 2.06 	& 6.72 	&& 10.94$\times$ 	& 6.29$\times$ 	& 9.52$\times$ 	&& 5.96$\times$\\
			neptune 	& dragon2 	& 6.2M 	&& 44.53 	& 139.26 	& 183.78 	&& 32.65 && 4.02 	& 2.04 	& 6.06 	&& 11.08$\times$ 	& 68.26$\times$ 	& 30.33$\times$ 	&& \textcolor{blue}{\textbf{8.12$\times$}}\\
			\rowcolor{Gray}
			ganesha 	& dragon2 	& 6.5M 	&& 35.44 	& 134.41 	& 169.85 	&& 14.90 && 3.62 	& 1.72 	& 5.34 	&& 9.80$\times$ 	& 78.15$\times$ 	& 31.83$\times$ 	&& 4.12$\times$\\
			ganesha 	& horse 	& 6.5M 	&& 35.22 	& 130.04 	& 165.26 	&& 13.46 && 3.48 	& 1.67 	& 5.15 	&& 10.12$\times$ 	& 77.82$\times$ 	& 32.08$\times$ 	&& 3.87$\times$\\
			\rowcolor{Gray}
			neptune 	& raptor 	& 6.0M 	&& 37.20 	& 138.95 	& 176.15 	&& 20.62 && 3.37 	& 1.70 	& 5.08 	&& 11.03$\times$ 	& 81.64$\times$ 	& 34.71$\times$ 	&& 6.11$\times$\\
			horse       & neptune 	& 6.2M 	&& 38.65 	& 11.38 	& 50.03 	&& 22.07 && 3.43 	& 1.63 	& 5.06 	&& 11.28$\times$ 	& 6.99$\times$ 	& 9.90$\times$ 	&& 6.44$\times$\\
			\rowcolor{Gray}
			dragon1 	& ganesha 	& 5.6M 	&& 32.93 	& 119.43 	& 152.35 	&& 11.59 && 3.47 	& 1.39 	& 4.86 	&& 9.50$\times$ 	& \textcolor{blue}{\textbf{85.79$\times$}} 	& 31.37$\times$ 	&& \textcolor{red}{\textbf{3.35$\times$}}\\
			ganesha 	& raptor 	& 6.3M 	&& 29.68 	& 125.36 	& 155.04 	&& 11.87 && 3.18 	& 1.61 	& 4.79 	&& \textcolor{red}{\textbf{9.35$\times$}} 	& 77.82$\times$ 	& 32.39$\times$ 	&& 3.74$\times$\\
			\rowcolor{Gray}
			dragon1 	& neptune 	& 5.3M 	&& 32.91 	& 8.99      & 41.90 	&& 18.55 && 2.76 	& 1.40 	& 4.15 	&& \textcolor{blue}{\textbf{11.93$\times$}} 	& 6.44$\times$ 	& 10.09$\times$ 	&& 6.72$\times$\\
			horse       & dragon2 	& 9.4M 	&& 27.28 	& 88.21 	& 115.49 	&& 10.44 && 2.36 	& 1.11 	& 3.47 	&& 11.56$\times$ 	& 79.40$\times$ 	& 33.26$\times$ 	&& 4.42$\times$\\
			\rowcolor{Gray}
			raptor      & dragon2 	& 4.2M 	&& 25.96 	& 87.85 	& 113.80 	&& 10.54 && 2.24 	& 1.05 	& 3.29 	&& 11.57$\times$ 	& 84.06$\times$ 	& 34.60$\times$ 	&& 4.70$\times$\\
			horse       & raptor 	& 4.2M 	&& 24.25 	& 80.78 	& 105.03 	&& 8.34 && 2.08 	& 0.98 	& 3.07 	&& 11.63$\times$ 	& 82.18$\times$ 	& 34.24$\times$ 	&& 4.00$\times$\\
			\rowcolor{Gray}
			dragon1 	& dragon2 	& 3.5M 	&& 22.22 	& 69.96 	& 92.17 	&& 8.33 	&& 2.01 	& 0.86 	& 2.88 	&& 11.04$\times$ 	& 80.97$\times$ 	& 32.04$\times$ 	&& 4.14$\times$\\
			dragon1 	& horse 	& 3.5M 	&& 22.06 	& 5.27      & 27.33 	&& 7.75 	&& 1.90 	& 0.83 	& 2.73 	&& 11.61$\times$ 	& 6.35$\times$ 	& 10.01$\times$ 	&& 4.08$\times$\\
			\rowcolor{Gray}
			dragon1 	& raptor 	& 3.3M 	&& 16.92 	& 63.55 	& 80.47 	&& 6.37 	&& 1.54 	& 0.75 	& 2.29 	&& 10.97$\times$ 	& 84.96$\times$ 	& \textcolor{blue}{\textbf{35.13$\times$}} 	&& 4.13$\times$\\
			\bottomrule\\
		\end{tabular}%
	}
	\caption{Comparative analysis of our Boolean algorithm w.r.t. any combination of the huge models in Figure~\ref{fig:bigmeshes}. For each test we report the name of the two objects and their cumulative size, in millions of triangles. With libigl we denote the best performing implementation of~\cite{zhou2016mesh}, which endows a well crafted parallelization that fully exploits the recently improved thread safety of the lazy rational kernel in CGAL 5.4. FA is the author's reference implementation of the Fast Arrangement algorithm~\cite{CLSA20_arrangement}. Running times are in seconds. All tested methods perform equally well on any Boolean operator, we therefore restricted experiments to Boolean unions only. The rightmost sections of the table, denoted with $\times$, report our speedups. On average, we are almost 25$\times$ faster than libigl and our arrangement is more than 5$\times$ faster than~\cite{CLSA20_arrangement}. Best and worst speedups for each algorithmic step are highlighted in bold blue and red, respectively. All tests were performed on a Mac Book M1 Pro with 8 performance cores and 32GB of RAM.}
	\label{tab:bigmeshes}
\end{table*}

\paragraph{ARAP deformation.} In this demo, two triangle meshes are interactively deformed using ARAP~\cite{sorkine2007rigid}. At first, the user prescribes an arbitrary number of handles on both shapes with mouse clicks. Then, matrices are factorized and the interactive session starts. During interaction the user can select any handle from both shapes and freely move it in space. The program updates the input meshes with ARAP and immediately performs a Boolean between them (Figure~\ref{fig:arap_demo}). Users can also interactively change the Boolean operator in real time. This demo is more challenging than the previous one, since all inputs to the Boolean algorithm undergo non trivial dynamic changes at interaction time, and since the cost of ARAP is roughly equivalent to the cost of Booleans.
To obtain maximum efficiency 
we execute the ARAP and Boolean steps in parallel on the same work stealing scheduler, using double buffering to batch operations within the stages. This allows us to hide latency as much as possible. In all our tests we always performed four iterations of ARAP, which are typically sufficient to obtain visually pleasant deformations. As can be noticed from the attached video, deformations appear quite smooth up to scenes containing 100K triangles, and progressively lag for bigger scenes containing 150K and 200K triangles. Detailed numbers on our running times are reported in Table~\ref{tab:fps_arap}.

It should be noted that ARAP deformation offers no guarantees, and under extreme handle displacements may occasionally introduce self intersections that violate the input requirements of our method and may possibly spoil the in/out classification system. In our experiments we observed that artifacts of this kind occasionally arose when rotation matrices where computed with~\cite{Zhang:Fast:2021}. Using 
Eigen SVD~\shortcite{eigenweb} for the computation of rotation matrices solved all our issues, although the algorithm becomes a bit slower (the local step was $1.5\times$ faster using~\cite{Zhang:Fast:2021}). 

\begin{figure*}[h]
	\centering
	\includegraphics[width=\linewidth]{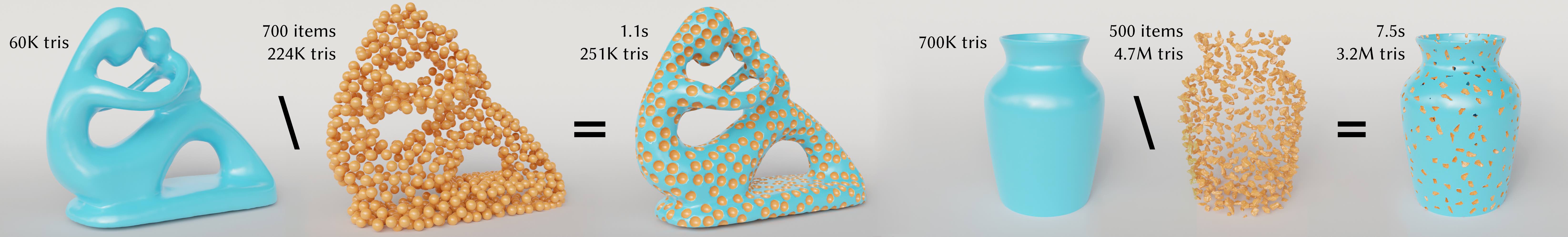}
	\caption{Our method scales optimally to variadic Booleans involving hundreds of shapes. We executed the operations in the figure in two different ways: passing in input each decorative element separately, and merging all decorative elements into a single input mesh. In both cases the running time was the same, thus no computational overhead was introduced for the increased number of input shapes.}
	\label{fig:variadic}
\end{figure*}

\subsection{Large Scale Benchmark}
\label{sec:benchmark}

We considered the popular Thingi10K~\cite{thingi10k} dataset to perform a large scale benchmark, comparing our performances with~\cite{zhou2016mesh}. We compared to the recently released implementation by the authors that includes significant parallelization efforts.
Both methods require the input meshes to be manifold watertight, therefore we extracted 7628 \textit{clean} meshes from the version of the database released by the authors of~\cite{hu2018tetwild}. We halved these meshes in two groups of 3814 elements each and randomly combined them together to perform a Boolean operation. To make sure that the shapes actually intersect with each other we normalized their bounding box and centered them in the origin.
Since all methods are exact, they are also guaranteed to produce exactly the same output topology. We exploited this property to validate our algorithm, verifying that indeed the number of connected components and Euler characteristic was the same for each output mesh.
We used a machine equipped with 12 cores and 128GB of RAM as testing hardware. On this machine, \cite{zhou2016mesh} can process all 3814 Booleans in 28.3 minutes, whereas our software terminated the same task in 4.5 minutes, also being faster in the $100\%$ of the cases. Note that the recent update of libigl's implementation is a considerable speed up compared to the one used in the original paper, bringing the times from 80 minutes of the original publication to 28.3 minutes for their current version.
Overall, both methods spent most of the computation in the arrangement part: 22.5 minutes~\cite{zhou2016mesh} and 4 minutes ours ($5.5\times$). For the Boolean part, \cite{zhou2016mesh} spent 5.8 minutes while our tool completed in 0.47 minutes ($12.2\times$). Remarkably, for the inside/outside classification our method based on ray casting was up to $101\times$ faster than the one of~\cite{zhou2016mesh} in the best case ($2.38\times$ in the worst case, $11.34\times$ on average).


\subsection{Processing of Huge Meshes}
\label{sec:hugemeshes}

Thingi10K is mostly populated by medium, small and very small meshes. We complement the large scale benchmark with a smaller test that focuses on high resolution meshes containing millions of triangles. While such big meshes are far from being suitable for interactive usage, this is the common size that can be found in production in many industries and is therefore practically relevant. We considered the eight high resolution meshes shown in Figure~\ref{fig:bigmeshes}, whose polygon count is in between 1.3 and 14.4 millions of triangles. We performed a Boolean operation for any possible pair of meshes, measuring running times of our tool, with~\cite{zhou2016mesh} and with~\cite{CLSA20_arrangement}, the latter only for the arrangement part. As shown in Table~\ref{tab:bigmeshes}, all methods scale well on very large datasets, mostly maintaining a stable ratio between their running times. Our method consistently operates one order of magnitude faster than~\cite{zhou2016mesh} for the arrangement part, where it is also $5\times$ faster than \cite{CLSA20_arrangement} on average. The highest variability occurs in the Boolean part, where the speedup compared  to~\cite{zhou2016mesh} oscillates around $80\times$ (in 70\% of the cases) and around $6\times$ (in the remaining cases). All in all, our average speedup w.r.t. to the whole Boolean pipeline of~\cite{zhou2016mesh} is approximately $25\times$. It is interesting to notice that for most of these tests the bottleneck for~\cite{zhou2016mesh} was the second step of the Boolean pipeline, which took most of the running time. This never happened for the small models in Thingi10K, where the Boolean part was almost negligible compared to the arrangement one. We conjecture that the topological propagation used in~\cite{zhou2016mesh} does not scale well on very big meshes, while our novel approach based on ray casting  remains efficient at all resolutions ($53\times$ faster on average) and its computation time never exceeded the cost of the arrangement part in any of our experiments.

%

\subsection{Variadic Booleans}
\label{sec:variadic}
Not only the mesh size but also the number of input objects affects the performances of a Boolean algorithm, especially for the second part of the pipeline, where the inside/outside relationships must be devised for each input shape. We evaluated the scalability of our method with respect to the number of input meshes involved in a Boolean operation, considering the subtraction between a base mesh $A$ and a large number of non intersecting small decorative elements $B_1 \cup B_2 \cup \dots \cup B_n$,  positioned on its surface with Poisson sampling~\cite{corsini2012efficient}. To isolate the impact of the number of inputs, we performed this experiment twice. The first time we consider the case of $A \setminus \left\lbrace B_1 \cup B_2 \cup \dots \cup B_n \right\rbrace$ as a variadic Boolean, that is, providing in input $n+1$ separate meshes. The second time, we merge $B_1 \cup B_2 \cup \dots \cup B_n$ in a single mesh $\dot{B}$ and then perform a classical pairwise operation $A \setminus \dot{B}$. Since decorations do not interfere with each other, the arrangement and the output result of these operations is identical.
Figure~\ref{fig:variadic} shows two results obtained with our tool. In the first one, we carved the Fertility statue (60K triangles) with 700 little spheres (each one counting 320 triangles, for 224K triangles overall). Our algorithm was able to correctly compute the same result with both approaches in 1.1 seconds each time, thus introducing no measurable overhead for the increased number of input meshes. In the second experiment we subtracted from a vase (700K triangles) an assembly of 500 little stones of 10 different types (from 5 to 16K triangles each, 4.7M triangles overall) randomly oriented and positioned on the surface of the vase. Once again, our algorithm computed both the variadic and the pairwise Boolean in almost the same time (7.49 and 7.59 seconds, respectively). This is possible thanks to our inside/outside labeling based on ray casting, which shoots a ray for each surface patch regardless of the number of input shapes. Prior methods based on topological flooding do not exhibit the same desirable property and tend to introduce unnecessary overhead when the number of inputs grows. For example, on the first experiment the method of~\cite{zhou2016mesh} had a slowdown factor of more than $5\times$ between the two runs, completing the Booleans in 6.8 and 38 seconds, respectively. On the second experiment their running times were 61.01  and 170.93 seconds (2.8$\times$).

\section{Conclusion}
We have presented a novel pipeline for the computation of topologically exact mesh Booleans. Our main technical contributions amount to an amelioration of the arrangement algorithm in~\cite{CLSA20_arrangement}, and to a novel inside/outside classification system based on exact ray casting. As shown in our experiments both contributions are significantly faster than prior art, by at least one order of magnitude overall. Thanks to this speedup we could implement interactive applications that couple basic geometry processing tasks with real-time Booleans. This is the first time robust Booleans and interactive tools are coupled together for real sized meshes. To this end, we expected the community of digital artists and context creators to readily adopt our tools, and we are curious to see what they will be able to create with it.

\subsection{Limitations and Future Works}
\label{sec:future_works}
Our system is currently limited in two aspects: inability to achieve interactive frame rates on very high resolution meshes (e.g. more than 200K triangles) and inability to robustly perform cascaded Boolean operations.

\paragraph{Scalability.} The lowest hanging fruit to improve on the scalability of our method in interactive mode is to cache partially evaluated computations between frames. As discussed in Section~\ref{sec:interapp}, the algorithm is at the moment not designed to exploit temporal coherency and naively computes each Boolean from scratch, which is of course not optimal. Things like acceleration data structures used to detect intersection and perform the ray casting could be created once and minimally updated at each frame, greatly reducing the computational cost. Also adjacency data is now recomputed from scratch each frame, while they can be cache in most cases. 

\paragraph{Cascaded Booleans.} Our interactive tools are currently limited to CSG-like applications and other applications where each frame can be generated independently from the previous ones.  We currently do not support \emph{cascaded} Booleans, where the inputs are the result of a previous Boolean operation.
While in our tool it is technically possible to cascade Booleans, no guarantees on the result can be given because after each frame we snap exact coordinates to floating points, possibly introducing small mesh defects that may spoil subsequent operations. As mentioned at the end of Section~\ref{sec:interapp} similar problems may also occur when Booleans are coupled with geometry processing tasks that do not guarantee the absence of self-compenetration, like ARAP. This issue was solved in \cite{diazzi2021convex} by implicitly repairing the input during the process, though the need to construct an intermediate volume necessarily introduces a slowdown. Solving the snap rounding problem is the ultimate solution for all these issues, but this is a remarkably difficult problem as previously discussed. Avoiding the snap rounding step and propagating the implicit points naively is not possible either, because the repeated composition of implicit points would soon lead to complex polynomial expressions for which the arithmetic filtering fails in virtually all cases, introducing major slowdowns, and eventually running out of memory. On the one hand, we argue that such a repeated composition is not strictly necessary for cascading, because the output of a Boolean operation is always a subset of the input (ground truth) primitives, but we lack a proper mechanism to \emph{update} the definition of implicit intersection points so as to avoid second level constructions. In the context of plane-based representations a similar idea was recently explored in~\cite{nehring2021fast}. Attempting to realize cascading for the Indirect Predicates~\cite{attene2020indirect} is an interesting direction for future research.

\bibliographystyle{ACM-Reference-Format}
\bibliography{biblio}

\end{document}